Nicholas James Rowe

# Applications of band-limited extrapolation to forecasting of weather and financial time series

School of Electrical Engineering, Computing and Mathematical Sciences, Curtin University

Abstract

This paper describes the practical application of causal extrapolation of sequences for the purpose of forecasting. The methods and proofs have been applied to simulations to measure the range which data can be accurately extrapolated. Real world data from the Australian Stock exchange and the Australian Bureau of Meteorology have been tested and compared with simple linear extrapolation of the same data. In a majority of the tested scenarios casual extrapolation has been proved to be the more effective forecaster.

## Section 1: Introduction

Forecasting signal and sequences with unknown underlying processes it a much sought-after thing. The Bureau of meteorology had over $387million in financial spending for 2017-2018 a portion of which is dedicated to climate and temperature forecasting using complex climate maps [2].While the ability understands the underlying process of stock and index funds is impossible. In this report we will attempt to forecast the outcomes of these two fields using the process of causal smoothing and extrapolation.

This paper applies the method of causal smoothing that have been detailed and mathematically proven in [1] and attempts to apply them in order to forecast the maximum price of stocks and index funds and the maximum daily temperatures of a weather station with no knowledge of the underlying process and compare it to simple linear extrapolation.

## Section 2: Theory and definitions

The primary source for the methods used within this paper are sourced from the paper "On Causal extrapolation of sequences with application to forecasting" By Nikolai Dokuchaev [1] . A brief summary of these definitions and methods will be listed below for completeness, the most import of which is the method by which a sequence of vectors may undergo causal approximation and extrapolation. For proof of these methods please see the original paper as this section is largely quotes of the previous.

### Definitions
We define the following

$$\mathrm{sinc}(x) = \frac{\sin(x)}{x}$$

$Z$ is the set of all integers, $\Omega \in (0, \pi)$ and is a positive real number $N$

s is an element of the set of all integers, q is an element of Z smaller then s or is negative infinity

$$s \in Z \text{ and } q \in \{ k \in Z : k < s \} \cup \{-\infty\}$$

T is the integer region between s and q or s and negative infinity if q is negative infinity.

$$\mathbf{T} = \{t \in Z: q \leq t \leq s\} \text{ if } q > -\infty \text{ and } \mathbf{T} = \{t \in Z: t \leq s\} \text{ if } q = -\infty$$

$$Z_n \text{ is the set of all integers } |k| \leq N$$

$l_r^+$ is the set of all sequences where x ∈t such that x(t)=0 for all negative integers of t

$$\text{If } x \in l_2, \text{ then } X|_T \text{ is defined as an element } L_2(T)$$

Tau is an element of the set of all integers and positive infinity, and theta is less than tau including negative infinity.

$$\text{Let } \tau \in Z \cup \{+\infty\} \text{ and } \theta < \tau; \text{ the case where } \theta = -\infty \text{ is not exluded.}$$

We denote by $l_2(\theta, \tau)$ the Hilbert space of complex valued sequences $\{x(t)\}_{t=\theta}^{\tau}$ such that $\|x\|_{l_2(\theta,\tau)} = (\sum_{k \in Z_n} |x(t)|^2)^{1/2} < +\infty$

$Y_n$ is the hilbert space of sequences $\{y_k\}_{k=-N}^{N} \subset$

C provided with the $l_2$ norm $\|y\|_{Y_n} = \left(\sum_{k \in Z_n} |y(t)|^2\right)^{\frac{1}{2}} < +\infty$

$l_2^{BL}$ is the set of z on $l_2$ such that the Z transform of z is 0 for $|\omega| > \Omega$. The inverse transform of this will be called band limited

$l_2^{BL}$ be the set of all mapping $x \in l_2$ such that $X(e^{iw}) \in L_2(-\pi, \pi)$ the $X(e^{iw}) = 0$ for $|\omega| > \Omega$ where $X = Zz$.

We will call the corresponding process z=Z⁻¹X band limited

Bn is the set of Z transforms of z(t) where q<t<s and are integers, such that there exists a sequence in Yn.

$$B_N \text{ is the set of } X \in L_2(\mathbf{T}) \text{ such that there exists a sequence}$$

$$\{y_k\}_{k=-N}^{N} \in Y_n \text{ such that } X(e^{i\omega}) = \sum_{k=-N}^{N} y_k e^{\frac{i\omega\pi}{\Omega}} I_{\{|\omega| \leq \Omega\}}$$

$$\text{where } I \text{ is the indicator function.}$$

Consider the hilbert space $X = l_2$ and $X_- = l_2(q, s)$.

Let $B_n$ be the subset of $X_-$ consisting of sequences $\{x(t)\}_{t \in T}$ where $x \in X$ are such that $x(t) = Z^{-1}X(t)$ for $t \in \mathbf{T}$ for some $X(e^{iw}) \in B_n$

A sequence going from s to negative infinity is left band limited if there exists a band limited sequence equal to it for t<=s.

We call a one sided sequence x*∈$l_2$(-∞,s) left band limited if there exists x∈$l_2^{BL}$ such that x(t)=x*(t) for t≤s.

We denote by $l_2^{LBL}$(-∞,s) the set of all these sequences x*, and we denote by $l_{2,N}^{LBL}$(-∞,s) the set of all sequences x* such that the corresponding extrapolation x belongs to $B_N$

## Theory

The goal of these experiments is to minimize the residual between raw data points z(t) and values generated from causal smoothing extrapolation points $\hat{x}(t)$.

$$\text{There exists an optimal solution } \hat{x} \text{ of the minimization problem}$$

$$\text{Minimise} \sum_{t=s}^{s+D} |\hat{x}(t) - z(t)|^2 \text{ over } \hat{x} \in B_N$$

This does not necessarily mean minimising the residual of the causal smoothing approximation as doing so will result in no extrapolated values.

If the sequence z(t) is left band limited where q<t<s, its extrapolation is uniquely defined by its history such that if s-q≤2N+1 then there exists a unique z-transformation that is an element of $B_N$.

"For $x_L \in l_2^{LBL}(-\infty, s)$, the extrapolation $\bar{x} \in l_2$ described is uniquely defined. If $s - q < 2N + 1$, then for any $X \in B_N$ there exists a unique $X \in B_N$ such that $z(t) = (Z^{-1}X)(t)$ for $t \in T$."

## Z-Transform to operator Q
This z-transform can be expressed as a sinc(x) function by the following method

"Let the operator $Q: Y_n \to B_n$ be defined as $\hat{x} = Qy = Z^{-1}\widehat{X}$

$$\widehat{X}(e^{i\omega}) = \sum_{k=Z_N} y_k e^{\frac{i\omega\pi}{\Omega}} I_{\{|\omega| \leq \Omega\}}$$

$$\widehat{X}(e^{i\omega}) = \frac{1}{2\pi} \int_{-\Omega}^{\Omega} (\sum_{k=Z_N} y_k e^{\frac{i\omega\pi}{\Omega}}) e^{i\omega t} d\omega = \frac{1}{2\pi} \sum_{k=Z_N} y_k \int_{-\Omega}^{\Omega} e^{\frac{i\omega\pi}{\Omega} + i\omega t} d\omega$$

$$\frac{1}{2\pi} \sum_{k=Z_N} y_k \frac{e^{ik\pi + i\Omega t} - e^{-ik\pi - i\Omega t}}{\frac{ik\pi}{\Omega} + it} d\omega = \frac{\Omega}{\pi} \sum_{k=Z_N} y_k \, sinc(k\pi + \Omega t) = Qy(t)$$

It follows that the $Q: Y_N \to B_N$ is actually defined as

$$\hat{x}(t) = (Qy)(t) \frac{\Omega}{\pi} \sum_{k=Z_N} y_k \, sinc(k\pi + \Omega t)$$

Consider the operator $Q^*: B_N \to Y_N$ being adjoing to the operator $Q: Y_N \to B_N$ i.e, such that

$$(Q^*X)_k = (Qy)(t) \frac{\Omega}{\pi} \sum_{k=Z_N} y_k \, sinc(k\pi + \Omega t) \, z(t)"$$

By the property of the sinc function, it follows that this convolution map continuously $l_2$ into $l_2$. Hence the operator $Q^*$ can be extended as a continuous linear operator $Q^*: X_- \to Y_N$.

## The Operator R
Consider the linear bounded non-negative definite Hermitian operator $R: Y_N \to Y_N$. defined as

$$R = Q^*Q$$

The operator $R: Y_N \to Y_N$ has a bounded inverse operator $R^{-1}: Y_N \to Y_N$

## The Operator Q⁺

$$\hat{x} = QR^{-1}Q^*z$$

$$\hat{x}(t) = \hat{x}(t, q, s) = \frac{\Omega}{\pi} \sum_{k=Z_N} \widehat{y_k} \, sinc(k\pi + \Omega t)$$

Here $\hat{y} = \{\widehat{y_k}\}_{k=-N}^{N}$ is defined as

$$\hat{y} = R^{-1}Qz$$

R can be represented via a matrix $= \{R_{km}\} \in C^{2N+1,2N+1}$ , where k,m=-N, -N+1 ,… N-1, N. In this setting,
$(Ry)_k = \sum_{k=-N}^{N} R_{km} y_m$ and the components of the matrix R are defined as

$$R_{km} = (\frac{\Omega}{\pi})^2 \sum_{t=q}^{s} sinc(m\pi + \Omega t) sinc(k\pi + \Omega t)$$

Respectively, the components of the vector $Q^*x = \{(Q^*x)_k\}_{k=-N}^{N}$ are defined as

$$(Q^*z)_k = \frac{\Omega}{\pi} \sum_{t=q}^{s} sinc(k\pi + \Omega t) z(t)$$

The process $\hat{x}$ represents the output of a linear causal smoothing filter. It can be noted that the operators R and Q have to be recalculated for each s, and values $\hat{x}(t) = \widehat{x_{q,s}}(t)$ calculated for the observations $\{x(t), s \leq t \leq q\}$ , can be different from values $\widehat{x_{q,s+\tau}}(t)$ calculated for the same t using the observations $\{x(t), s \leq t \leq q + \tau\}$, where $\tau > 0$ therefore , this filter is not time invariant.

*Tikhonov regulation*
This lets us consider a modification of the original optimization problem with penalty on the norm of the solution that restrains the norm of the solution. More precisely, let us consider the following problem;

$$Minimize \; ||\hat{x} - x||_{X_-}^2 + v||\hat{x}||_{l_2}^2 \; over \; \hat{x} = B_N$$

This has a unique solution as follows.

$$\hat{x} = Q R_v^{-1} Q^* z$$

where

$$R_v = R + vI$$

Where I is the identity matrix $R^{NxN}$.

As previously stated the above definitions and theories have been quoted from paper "On Causal extrapolation of sequences with application to forecasting" By Nikolai Dokuchaev [1] as it is the most comprehensive summary of the necessary parts of this process.

## Summary
Raw data z(t) must undergoes several operators in order to be transformed into ideal solution $\hat{x}(t)$. First the operator $Q^*$ will be applied in order to take z(t) from Bn to Yn.

$$(Q^*z)_k = \frac{\Omega}{\pi} \sum_{t=q}^{s} sinc(k\pi + \Omega t) z(t)$$

This is then applied to the inverse R matrix with Tikhonov regulation which will map it from Yn to Yn

$$R_{km} = (\frac{\Omega}{\pi})^2 \sum_{t=q}^{s} sinc(m\pi + \Omega t) sinc(k\pi + \Omega t)$$

With the addition of the Tikhonov regulation of vI .This is the equivalent to the Q⁺ operator and creates the sequence y$_k$. This then has the operator Q applied to bring it back to Bn as a causally smoothed version of itself.

$$\hat{x}(t) = (Qy)(t) \frac{\Omega}{\pi} \sum_{k=Z_N} y_k \, sinc(k\pi + \Omega t)$$

$\hat{x}(t)$ is a sequence that can be continued by inserting t>s into the above equation solving.

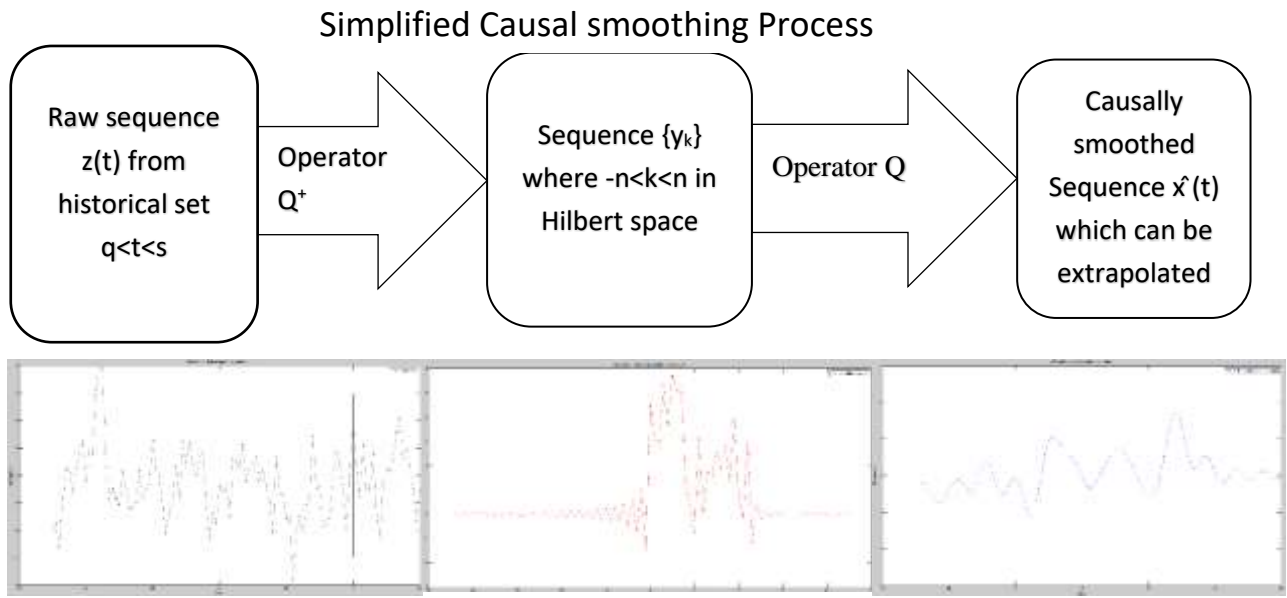

Image 2.1: Visual representation of the main process discussed in this paper. (Not to scale)

# Section 3

## Simulation: Causal Extrapolation of Monte Carol simulation

To test the extrapolating capacity of Left band limited causal smoothing a monte carol simulation was created as follows:

$$z(t) = A(t)z(t-1) + \sigma \eta(t) \; for \; t \in Z \; where \; Z \; is \; the \; set \; of \; real \; integers$$

Similar to the process used in ^^ however in this case z(t) is a process within $R^1$ where A(t) takes random values within a uniform distribution, $\eta(t)$ is standard gaussian white noise and

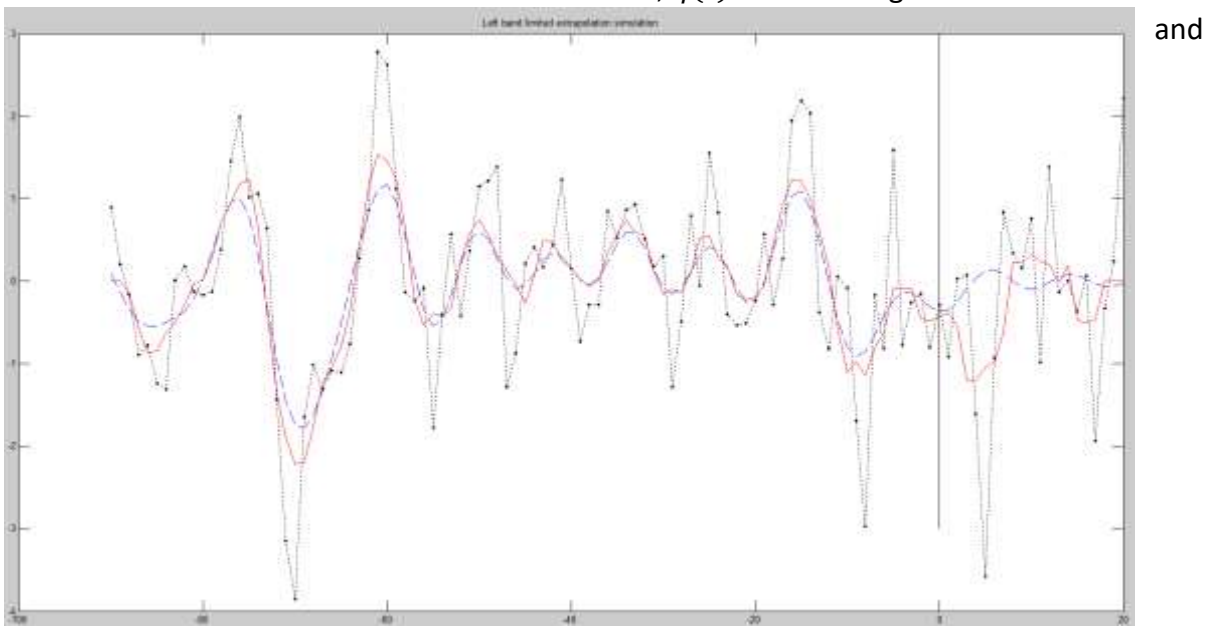

$\sigma$ is simply 1.

*Figure 3.1 Example of Monte Carol Simulation. Raw data: Black, Moving average: Red, Causal Extrapolation :Blue*

In order to get a better result, the process was stabilised using a 5-point moving average (MV), this was used as the values for $\{x(t), s \leq t \leq q\}$ explicitly q=-90 and s =0 and MV(t) =x(t) rather than the raw z(t) values.
The process is then as follows

$$\hat{x} = QR_\epsilon^{-1}Q^*x$$

$$(Q^*x)_k = \frac{\Omega}{\pi}\sum_{t=q}^{s} sinc\big((k\pi + \Omega t)/\pi\big)MV(t)$$

$$R_{km} = (\frac{\Omega}{\pi})^2 \sum_{t=q}^{s} sinc(m\pi + \Omega t)sinc(k\pi + \Omega t)$$

$$R_v = R + vI$$

$$y_k = R_v^{-1}Q^*x$$
$$\hat{x}(t) = Qy_k$$
$$\hat{x}(t) = \hat{x}(t,q,s) = \frac{\Omega}{\pi}\sum_{k=Z_N} \widehat{y_k}\, sinc\big((k\pi + \Omega t)/\pi\big)$$

t is then extended past s to extrapolate values past 0 in this case s+20. Multiple trial variables where you several test $\Omega = \pi/4$ and v=0.05 were found to be one of the more effective values and has been used for the remainder of this paper. After these values were settled on 10000 repetitions of the base simulation were evaluate to explore how far the extrapolation could reasonably be made. The residual for each of these points has been calculated using

$$\frac{\sum_{m=1}^{10000}|\widehat{x_m}(t) - z_m(t)|^2}{m} = R(t)$$

Where s<t<s+20 and m is the trial number

The following is the mean residual of 20 extrapolated points with N=45 for 2N+1 =91 points of data:

*Table 3.1 : Mean Residuals of extrapolated values based on position*

| Simulation: Mean residuals of casual extrapolation points of 10,000 monte carol simulation | | | | | | | | | | |
|---|---|---|---|---|---|---|---|---|---|---|
| R(t) | R(1) | R(2) | R(3) | R(4) | R(5) | R(6) | R(7) | R(8) | R(9) | R(10) |
|  | 0.7641 | 0.8382 | 0.9383 | 0.9655 | 0.9681 | 0.9843 | 0.9757 | 0.9621 | 0.9733 | 0.9656 |
|  | R(11) | R(12) | R(13) | R(14) | R(15) | R(16) | R(17) | R(18) | R(19) | R(20) |
|  | 0.9750 | 0.9685 | 0.9692 | 0.9727 | 0.9629 | 0.9562 | 0.9791 | 0.9710 | 0.9678 | 0.9717 |

The causal smoothing had a total residual of 61.8 or 0.6792 per point and the residual for extrapolated points was 19.03 or 0.9514 per point

The point residual drops off quickly but remains between .95 and 1 for all 20 points, this leaves us with the conclusion that short-term extrapolation is preferable but the possibility for longer periods exist, though this may be just a result of the simulation being stationary. Hence further tests try to only extrapolate 2-7 data points ahead of s

## Section 4

### Australian stock exchange and index fund testing

The data set for the testing of stock prices for the ASX200 and CBA were provided by Market index [3]consisting of the opening, high, low and close values of the two stocks between the 11th of October 2017 and the 10th of October 2018. These values were also confirmed with the Commonwealth Securities Limited quotes [4] of these data points. The ASX200 index fund stock was used as a model to test the appendixed code, once the code was working as intended the CBA stock was then substituted into code to obtain the results. This has been done to ensure show that the code works for any like data set and has not been tailor made for a specific set, thought that maybe a topic for future work in this field.

Three experiments we carried out for the data sets focusing on two variable t the day relative to the historical start and z(t) the maximum price the stock/index fund achieved that day. The experiment first a prediction of 20 days with N=45 for 91 days of history to make a prediction. Second a 5-day forecast repeated 32 times and compared to linear extrapolation, for a total of 160 days of extrapolated data. The third experiment was a 2-day forecast repeated 79 times and compared to linear extrapolation of the same thing.

### Experiment 1: Single 20 day forecast

The main problem with converting from a stationary process of a simulation at the origin to an actual data point with an unknown process and a larger magnitude is maintaining a consistent magnitude for the smoothed data. The pattern of the process is copied at a lower price and then drops to zero when attempting to extrapolate the data points as can be seen in the graphic below.

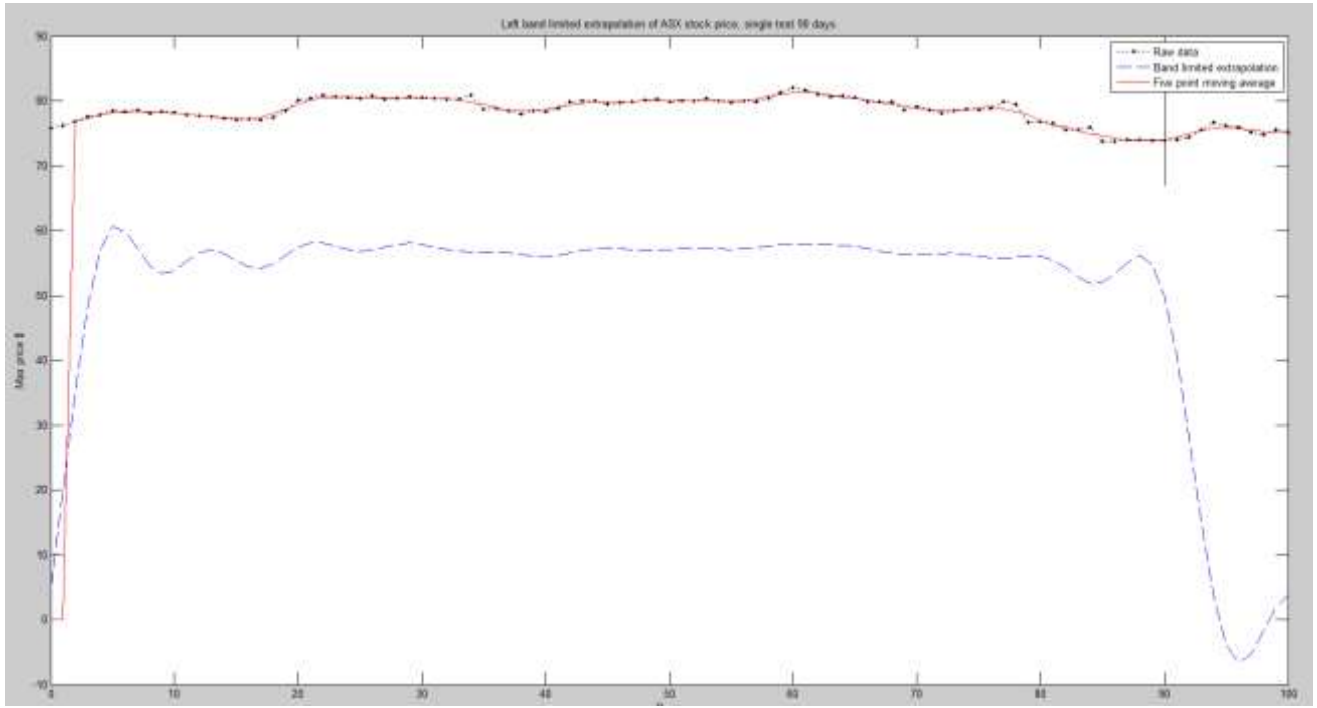

*Figure 4.1 Unfixed causal extrapolation for experiment 1*

| Experiment 1: Forecasting a single 20day forecast. Unfixed | | | |
|---|---|---|---|
| Historical Residual | Mean Historical Residual | Extrapolation Residual | Mean Extrapolation Residual |
| 2181.3 | $23.9701 | 643.5229 | $80.440 |
| Experiment 1: Forecasting a single 20day forecast. fixed | | | |
| Historical Residual | Historical Residual | Extrapolation Residual | Mean Extrapolation Residual |
| 216.95 | $2.3840 | 103.2814 | $5.1641 |

*Table 4.1 Experiment 1 residuals. Fixed and unfixed*

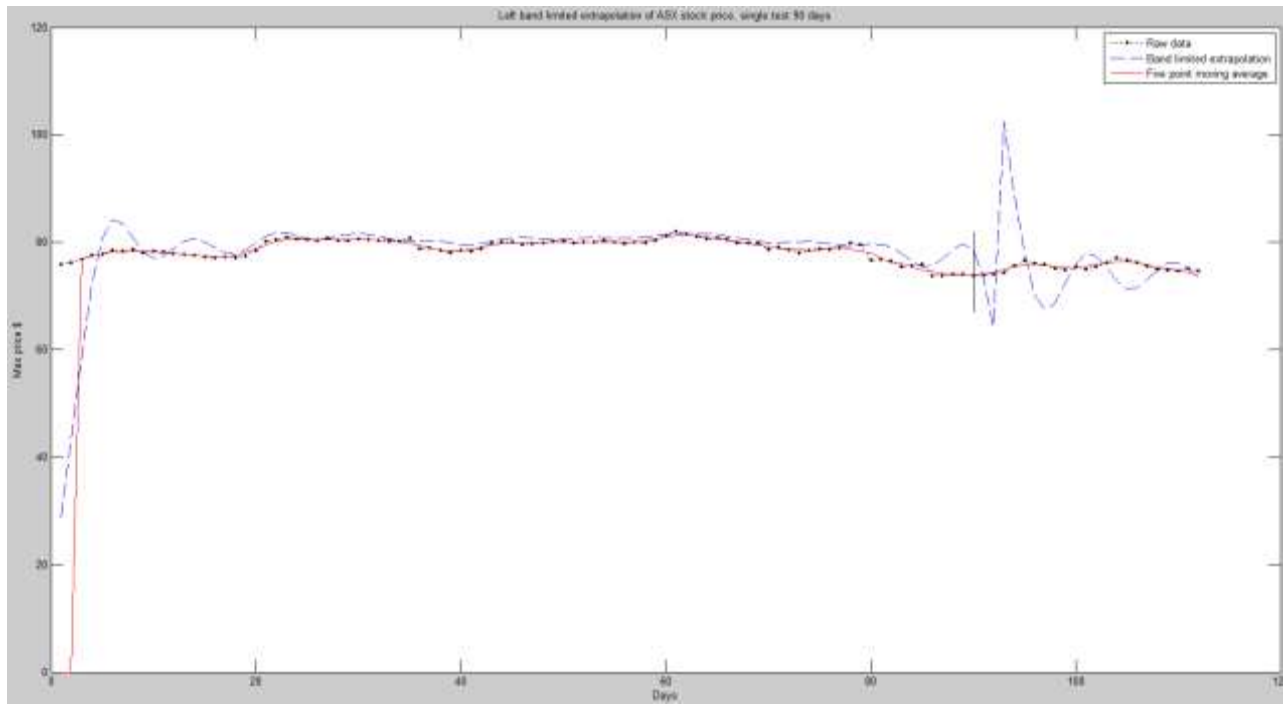

*Figure 4.2 Causal extrapolation for experiment 1, fixed.*

This drop of is a result of $\hat{x}(t)$ being created of components of $y_k$, naturally as the t exceeds the original x values the sinc(x) function would be more out of phase, hence returning a lower magnitude function with a similar pattern. The worst part of this is that the loss in magnitude of the function is far more pronounced then the pattern of the function, making the first 5 points unusable. As seen from the simulation, these are among the most accurate points. The current solution for this issue is overlapping data, where twice the extrapolated data points are created and the previous week fills in the subsequent repetitions first half. For example, in the second experiment the first extrapolation period creates 10 days (2 financial weeks) of extrapolated data, the last week is used as the first week for the second extrapolation period and so on for all 32 weeks of testing. All of these points are then boosted up by the last historical moving average point.

## Experiment 2 and Experiment 3 3 Results: 5 day forecast for 32 weeks & 2 day forecast repeated 78 times

Linear extrapolation has been chosen for comparison as it is a simple and reasonably effective way of forecasting financial data. The linear extrapolation is generated by creating a two-point line of best fit using 90 points of historical data and 5 points of historical data for experiment 2 and experiment 3 respectively. These were chosen for each as they performed best ASX200 trial set and there for the most competitive comparison.

*Linear extrapolation for A points historical data*

$$\hat{x}(t) = \frac{(z(t_0) - z(t_0 - A))}{A} (t) + (z(t_0) - \frac{(z(t_0) - z(t_0 - A))}{A}) t_0$$

In the second experiment the frequency casual smoothing extrapolation slightly out performed the linear extrapolation by about 16c per day over 160 days. This includes the rather wild projection of the first week caused by the drop off of $\hat{x}(t)$ between projection and extrapolation.

In the third experiment a linear extrapolation of the data outperformed the causal smoothing by approximately 4c per data point, including the initial drop off point.

These two experiments show that causal smoothing extrapolation is it least comparable if not better than standard linear extrapolation when forecasting financial data, this is in spite of the initial drop in $\hat{x}(t)$ and the resulting overlapping data.

Solving that problem by working out the drop of rate of $\hat{x}(t)$ and reversing it or by some other method of salvaging those first 1-5 data points should yield a result that is consistently more accurate then linear extrapolation of the data points.

| Experiment 2: 5 Day forecast | | | |
|---|---|---|---|
| Causal Smoothing Extrapolation Residual | Causal Smoothing Extrapolation Residual Mean | Linear Extrapolation residual (90 point) | Linear Extrapolation residual mean |
| $254.1679 | $1.5689 | $280.7319 | $1.7329 |
| Experiment 3: 2 Day forecast | | | |
| Causal Smoothing Extrapolation Residual | Causal Smoothing Extrapolation Residual Mean | Linear Extrapolation residual (5 point) | Linear Extrapolation residual mean |
| $147.8 | $0.9987 | $142.0 | $0.9597 |

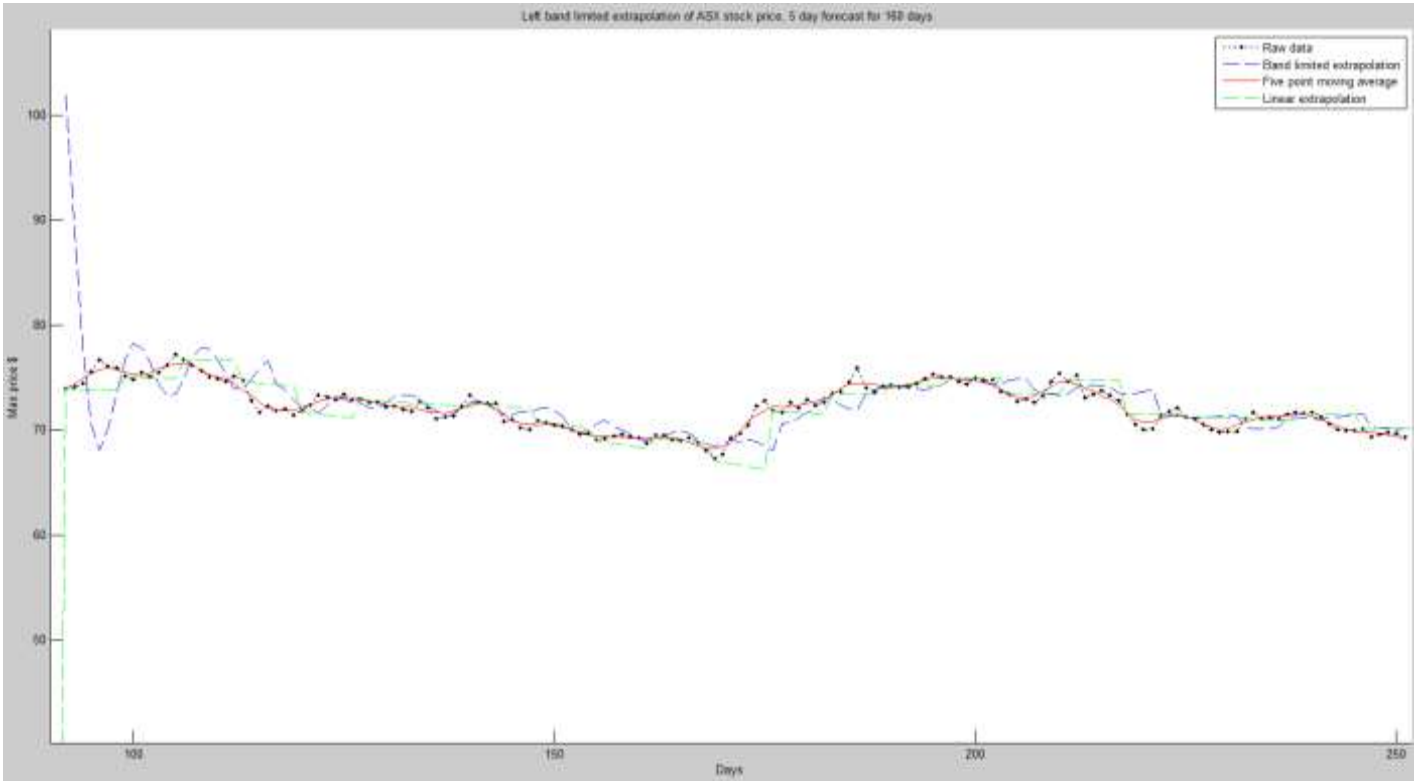

*Figure 4.3 Experiment 2, Causal extrapolation and linear extrapolations made every 5 days for 160days*

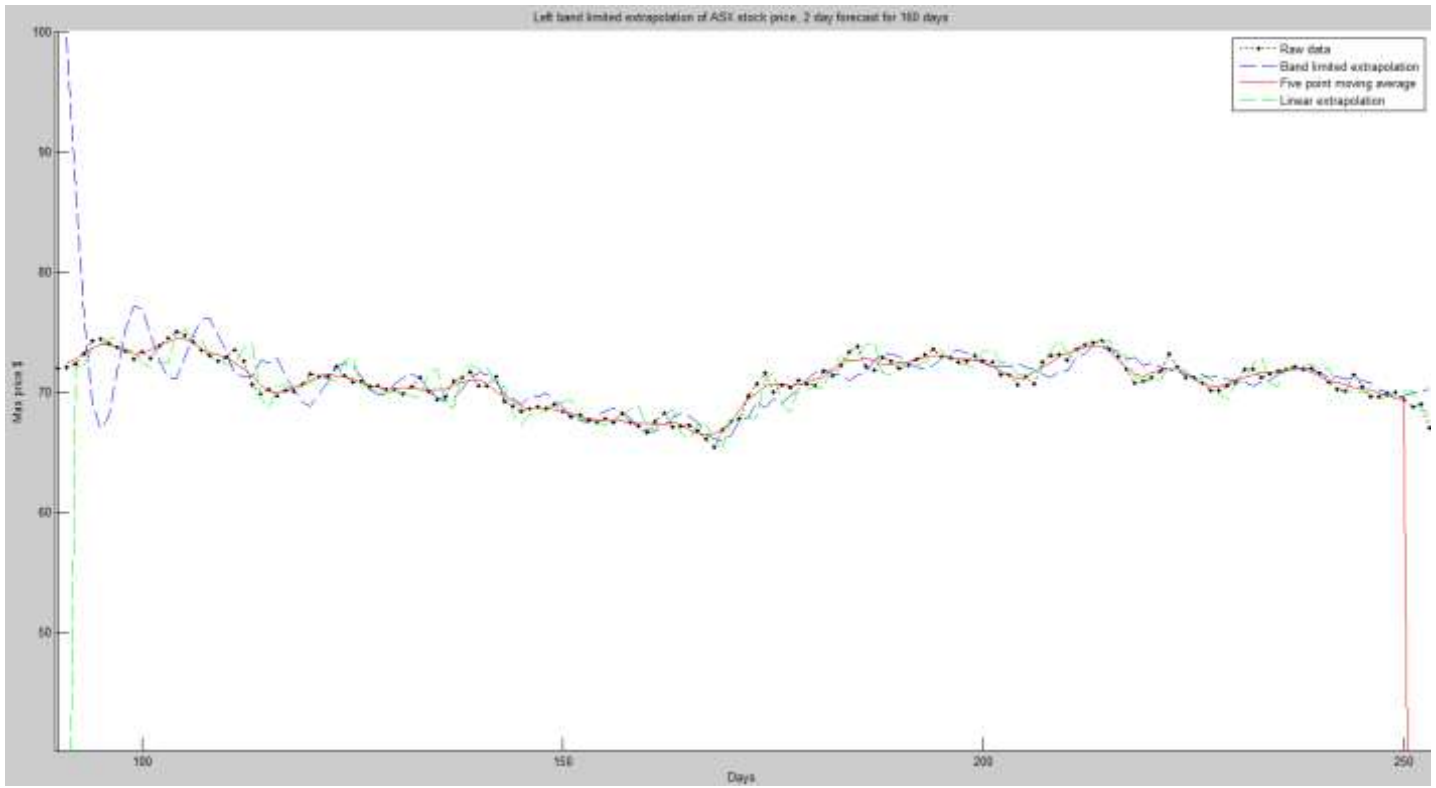

*Figure 4.4 Experiment 3, Causal extrapolation and linear extrapolation made every 2 days for 160 days*

# Section 5

## Bureau of Meteorology Maximum Temperature Extrapolation

Data from the Australian Bureau of meteorology was retrieved from the "Perth Metro" station located -31.9192 Latitude ,115.8728 Longitude at an altitude of 25.9 meters [5] [6]. Data ranged from the Stations first open in January first 1994 to the twenty fifth of October 2018. This data was selected as it had an estimated 100% completeness for Maximum air temperature data and over 2 decades of collected data as well as being the local weather station. For the purposes of testing two subsets of this data were used, the complete year or 2016 and 2017. Data mapped using Band limited smoothing suffered from a similar step-down problem as the Australian Stock exchange data, and required a similar step up using the mean residual for interpellated data and the mean of the last five moving average points. These adjustments were made done using the 2016 data as the control set and the 2017 data as the experiment data.

Three types of forecasts were made based of these data subsets, a once of 14-day forecast, a 7-day forecast repeated 38 times and a 2-day forecast repeated 133 times. All three use N=45 for 2N+1 = 91 for the historical set in order to make each of the forecast and have their data sets boosted up in the same way as the Australian Stock exchange Data, as well as the two moving data sets have the same overlapping data sets as the previous experiments. The Bureau of Meteorology was contacted in order to compare previous forecasts with these experiments forecast but the retrieval of the data exceeded the budget of this project, so the have instead been compared to linear extrapolation in the same way as the linear stock exchange data.

In both cases the Causal smoothing extrapolation outperformed the linear extrapolation in maximum temperature forecasts by at least a full degree. This is most likely because radical changes in temperature can easily "lead" linear extrapolation in the wrong direction creating massive residual for several extrapolation periods.

| Experiment 1: Single 14 Day forecast | | | |
|---|---|---|---|
| Extrapolation Residual | Average Extrapolation Residual | | |
| 35.3468 | 2.525 Degrees Celsius | | |
| Experiment 2: 7 Day forecasts repeated 38 times | | | |
| Causal Extrapolation Residual | Average Casual Extrapolation Residual | Linear Extrapolation residual (90 point) | Linear Extrapolation residual mean |
| 771.8907 | 2.7766 Degrees Celsius | 1019 | 3.816 Degrees Celsius |
| Experiment 3: 2 Day forecasts repeated 133 times | | | |
| Causal Extrapolation Residual | Average Casual Extrapolation Residual | Linear Extrapolation residual (90 point) | Linear Extrapolation residual mean |
| 492.2187 | 1.914 Degrees Celsius | 852.4756 | 3.343 Degrees Celsius |

*Table 5.1 Experimental Residuals for Causal and Linear Extrapolation*

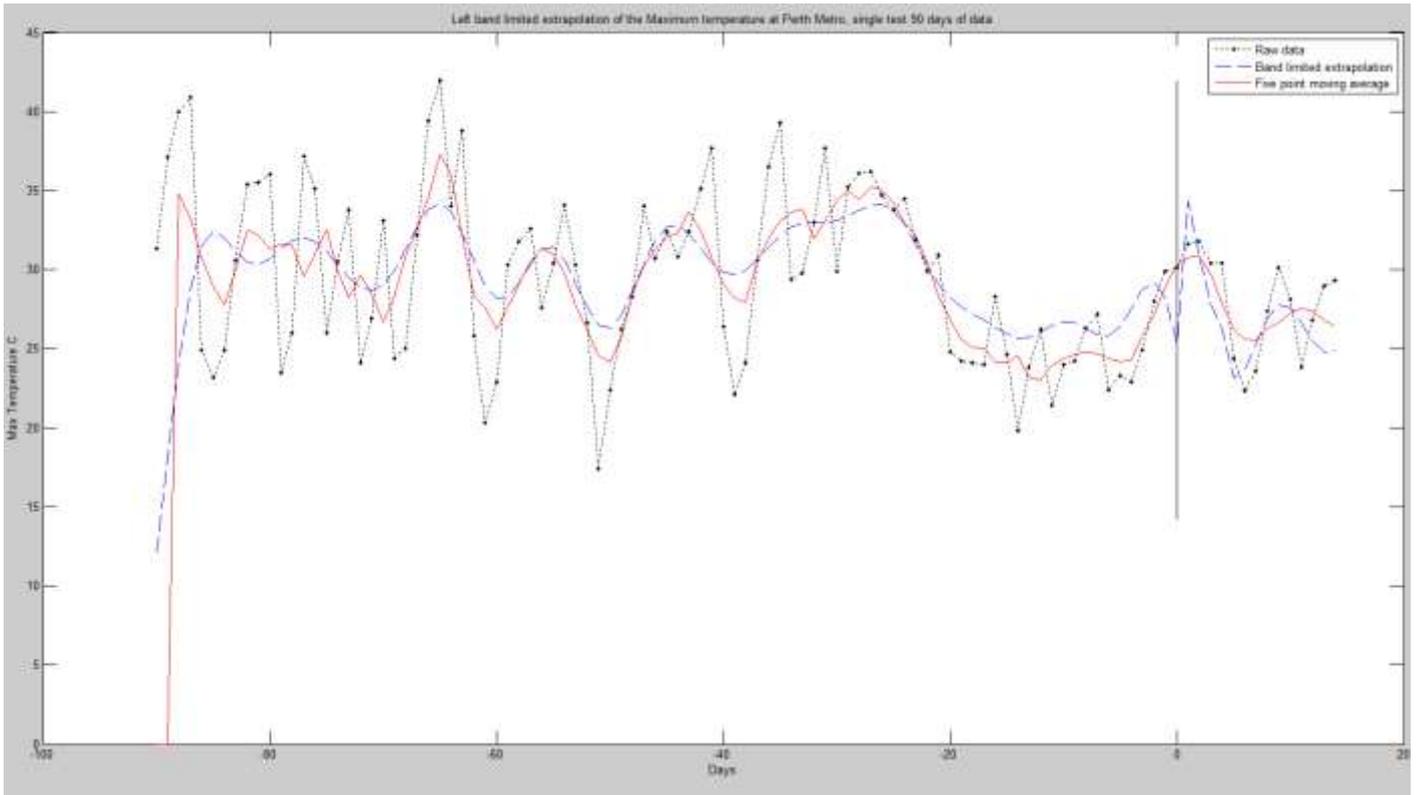

*Figure 5.1 Experiment 1 Causal extrapolation for a single 14 day forecast*

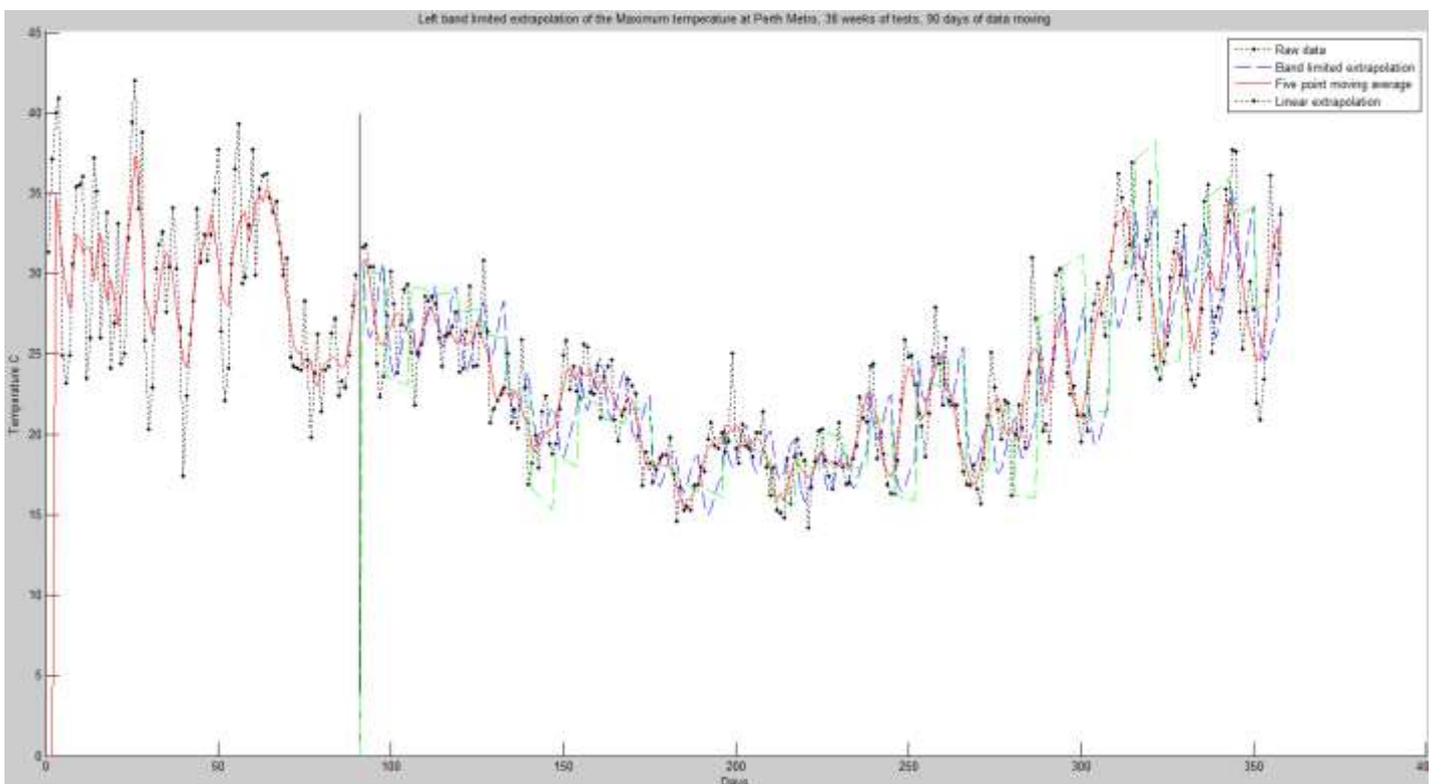

*Figure 5.2 Experiment 2 Causal and linear extrapolation made every 7 days for 38 weeks*

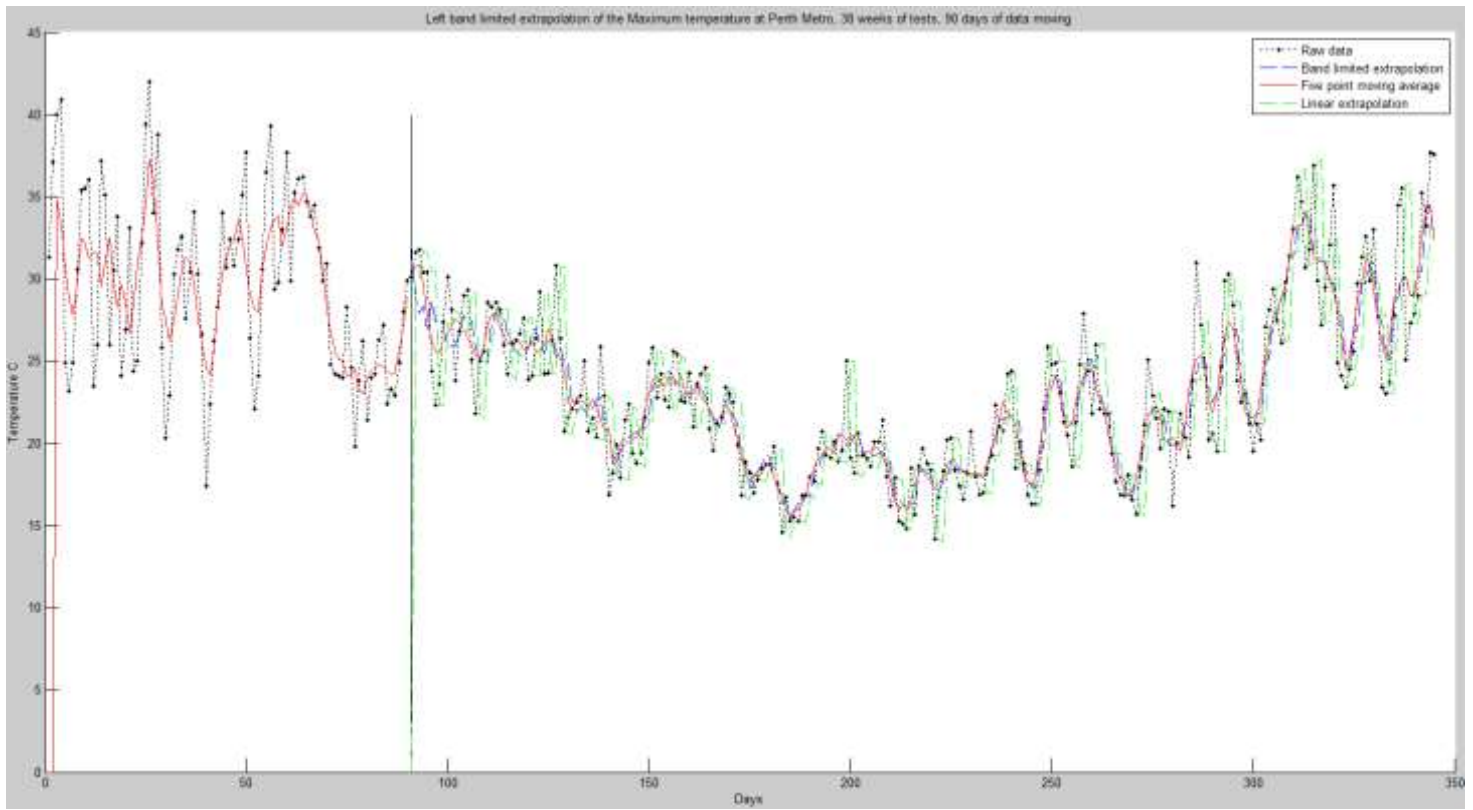

*Figure 5.3 Experiment 3 Causal and linear extrapolation made every 2 days for 38 weeks*

## Section 6: Summary and Conclusions

There is evidence that causal smoothing extrapolation can out perform linear extrapolation in cases where the underlying processes unknown, especially for data that changes gradient sharply. The Australian Stock exchange experiments showed causal smoothing extrapolation yielding forecasts with residuals $1.5689 over 5-day and $0.9987 over 2-day, compared with $1.7329 and $0.9597 for linear forecasting respectively. This demonstrates causal smoothing extrapolation maybe more effective for longer term forecasts then linear extrapolation but not for tomorrows forecast. In the case of the Bureau of Meteorology maximum temperature forecasts casual smoothing extrapolation outperformed linear extrapolation by a degree or more in all cases with 2.7766 Degrees Celsius and 1.914 Degrees Celsius for 7- and 2-day forecasts as compared to 3.816 Degrees Celsius and 3.343 Degrees Celsius for linear extrapolation, it is clearly a more reliable prediction method. The cause for this lack of accuracy in short term extrapolation is the loss in magnitude that occurs for causal extrapolation between smoothing data and extrapolating it. Summing the residuals found in section one we can find that the first two data points are have a residual that is 15.8% smaller then the following two, and the first 5 data point have a residual 7.95% of the following 5. It is not a great leap in logic to hypothesis that by solving this data loss between mapping and extrapolation, causal smoothing will be found to be the more accurate method of forecasting data with unknown process. Following that, the next step for this methodology will be to compare this spline extrapolation of real data sets and other more complex methods of forecasting real world processes.

# Section 7: References

# Section 8: Appendix

## Appendix 8.1 Monte Carol Extrapolation

```
function [ A,u,z ] = carlosim(N,o,z0,v )
% Takes Numbers of points N to create monte carol simulation of points over
% time. A(t) is a matrix of random uniform distribution. u is Gaussian
% white noise generated from a random normal distribution.
%Z(t)=A(t)Z(t-1)+o*u(t)
t = 2:1:2*N+1;
z = zeros(v,2*N+1);
% Number of colums depends on dimensions
u = random('Normal',0,1,v,1,2*N+1);
% Upper value depends on dimensions 1/v dimensions so 1D = 1 3D = 1/3
A = random('Uniform',0,1/v,v,v,2*N+1);
z(:,1)=A(:,:,1)*transpose(z0)+o*u(:,:,1);
for i=t;
    z(:,i)=A(:,:,i)*z(:,i-1)+o*u(:,:,i);
end

%Start by creating Monte-Carlo random numbers 1-D
%Z(t)=A(t)Z(t-1)+o*u(t)
%Input carlosim(N,o,z0,v) for N for 2N+1 points, z0 initial points, v
%dimensions, o>0 constant.
%Out puts A = v*v*N, u=v*1*N, z=v*1*N where z is the x(t) values
BLEP=zeros(22,1);
```

## Appendix 8.2 Band Limited Extrapolation of Monte Carol Simulation

```
[A,u,z]=carlosim(55,1,1,1);
X=pi;
N=45;
n=-N:1:N;
M=2*N+1;
%Currently unused
s=0;
q=-90;
ts=q:1:0;
%Moving average
MV=zeros(length(z),1);
MV(1)=(z(1)+z(2)+z(3)+z(4)+z(5))/5;
MV(1)=MV(2);
MV(1)=MV(3);
MV(length(z)-1)=(z(length(z)-1)+z(length(z)-2)+z(length(z)-3)+z(length(z)-4)+z(length(z)))/5;
MV(length(z)-1)=MV(length(z)-2);
MV(length(z)-1)=MV(length(z)-3);
for l=3:(length(z)-3)
    MV(l)=(z(l)+z(l+1)+z(l+2)+z(l-2)+z(l-1))/5;
end
%Process Q*R*Q+*x(t)
%Assign omega=g
g=pi/4;
%Creates list of Q+*x(t)
QX=zeros(M,1);
for k=-N:1:N
    QX(k+N+1)=0;
    for t=-90:1:0
        QX(k+N+1)=QX(k+N+1)+(g/pi)*sinc((k*pi+g*t)/X)*MV(t-ts(1)+1);
```

```matlab
        end
    end
%Creates matrix Rkm
R=zeros(M,M);
for k=-N:1:N
    for m=-N:1:N
        for t=ts

R(k+N+1,m+N+1)=R(k+N+1,m+N+1)+(g/pi)^2*sinc((k*pi+g*t)/X)*sinc((m*pi+g*t)/X);
        end
    end
end
%where v is epsilon
v=0.05;
RI= R+eye(M)*v;
y=inv(RI)*QX;
length(y)
%Creates list of Q sum of all Yk for each t
Q=zeros(111,1);
for t=-90:1:20
    for k=-N:1:N
        Q(t-ts(1)+1)=Q(t-ts(1)+1)+y(k+N+1)*(g/pi)*sinc((k*pi+g*t)/X);
    end
end
%Calulating residual
t=-90:1:0;
BLR=0;
for i=t
    BLR=BLR+sqrt((z(i+91)-Q(i+91))^2);
end
%Extrapolation
BLRE=0;
for i=1:1:20
    BLRE=BLRE+sqrt((z(i+91)-Q(i+91))^2);
end
for jj=1:1:20
    BLEP(jj)=BLEP(jj)+sqrt((z(jj+91)-Q(jj+91))^2);
end
BLEP(21)=BLEP(21)+BLR;
BLEP(22)=BLEP(22)+BLRE;
time = ii/10000

BLEP
BLEP/10000

length(z),length(Q),length(MV)
plot(-90:1:20,z(:),'k.:',-90:1:20,Q(:),'b--',-90:1:20,MV,'r')
title('Left band limited extrapolation simulation')
xlabel('Days')
ylabel('Max price $')
line([0,0],[-3,3],'Color',[0,0,0],'HandleVisibility','off')
hold all
plot(91:1:253,LE(1:163),'g--')
legend('Raw data','Band limited extrapolation','Five point moving average')
```

# Appendix 8.3 Band Limited Extrapolation of ASX: CBA Data single 20 day forecast

```matlab
%Start by importing raw data
A=xlsread('C:\Users\valough\Documents\MATLAB\CBA.AX.csv');
z=transpose(A(:,3));
X=pi;
N=45;
n=-N:1:N;
M=2*N+1;
%Currently unused
s=91;
q=0;
ts=q:1:s;
%Moving average
MV=zeros(length(z),1);
MV(1)=(z(1)+z(2)+z(3)+z(4)+z(5))/5;
MV(1)=MV(2);
MV(1)=MV(3);
MV(length(z)-1)=(z(length(z)-1)+z(length(z)-2)+z(length(z)-3)+z(length(z)-4)+z(length(z)))/5;
MV(length(z)-1)=MV(length(z)-2);
MV(length(z)-1)=MV(length(z)-3);
%Linear extrapolation
LE=zeros(10,1);
for tt=90:1:100
    LE(tt-89)=((z(tt)-z(tt-4))/5)*tt+(z(tt)-((z(tt)-z(tt-4))/5)*tt);

end
LE;
for l=3:(length(z)-3)
    MV(l)=(z(l)+z(l+1)+z(l+2)+z(l-2)+z(l-1))/5;
end
MA=mean(MV(5:length(MV)-5));
%Process Q*R*Q+*x(t)
%Assign omega=g
g=pi/4;
%Creates list of Q+*x(t)
QX=zeros(M,1);
for k=-N:1:N
    QX(k+N+1)=0;
    for t=ts;
        QX(k+N+1)=QX(k+N+1)+(g/pi)*sinc((k*pi+g*t)/X)*MV(t-ts(1)+1);
    end
end
%Creates matrix Rkm
R=zeros(M,M);
for k=-N:1:N
    for m=-N:1:N
        for t=ts

R(k+N+1,m+N+1)=R(k+N+1,m+N+1)+(g/pi)^2*sinc((k*pi+g*t)/X)*sinc((m*pi+g*t)/X);
        end
    end
end
%where v is epsilon
v=0.1;
RI= R+eye(M)*v;
y=inv(RI)*QX;
%Creates list of Q sum of all Yk for each t
Q=zeros(112,1);
for t=0:1:111
```

```matlab
        for k=-N:1:N
            Q(t-ts(1)+1)=Q(t-ts(1)+1)+y(k+N+1)*(g/pi)*sinc((k*pi+g*t)/X);
        end
end
%Calulating residual
t=1:1:112;
BLR=0;

D=Q(:);
%for i=1:1:length(Q)
    %for j=1:1:12
        %if zm(i)==j
        %   D(i)=D(i)+S(j);
        %        end

    % end
%end
for i=1:1:91
    BLR=BLR+sqrt((z(i)-(D(i)))^2);
end
BLR
D(1:92)=D(1:92)+BLR/91
D(93:112)=D(93:112)+MV(91)
BLRE=0;
for i=92:1:112
    BLRE=BLRE+sqrt((z(i)-(D(i)))^2);
end
BLRE
BLRE/20
mean(Q)
mean(z)
%88-94 problem zone
%Extrapolation
plot(t,z(1:112),'k.:',t,D(:),'b--',t,MV(1:112),'r')
title('Left band limited extrapolation of ASX stock price, single test 90 days')
legend('Raw data','Band limited extrapolation','Five point moving average')
xlabel('Days')
ylabel('Max price $')
line([90,90],[min(z),max(z)],'Color',[0,0,0])
hold all
```

## Appendix 8.4 Band Limited Extrapolation of ASX: CBA Data 5 Day forecasts

```matlab
%Start by importing raw data
A=xlsread('C:\Users\valough\Documents\MATLAB\CBA.AX.csv');
z=transpose(A(:,3));
N=45;
n=-N:1:N;
M=2*N+1;
X=pi;
%Currently unused
s=91;
q=1;
ts=q:1:s;
time=0
```

```matlab
%Moving average
MV=zeros(length(z),1);
MV(1)=(z(1)+z(2)+z(3)+z(4)+z(5))/5;
MV(1)=MV(2);
MV(1)=MV(3);
MV(length(z)-1)=(z(length(z)-1)+z(length(z)-2)+z(length(z)-3)+z(length(z)-
4)+z(length(z)))/5;
MV(length(z)-1)=MV(length(z)-2);
MV(length(z)-1)=MV(length(z)-3);
for l=3:(length(z)-3)
    MV(l)=(z(l)+z(l+1)+z(l+2)+z(l-2)+z(l-1))/5;
end
MA=mean(MV(5:length(MV)-5));
%Linear extrapolation
LE=zeros(10,1);
for tt=91:7:253
    mm=(z(tt)-z(tt-90))/90;
    cc=(z(tt)-((z(tt)-z(tt-90))/90)*tt);
    for it=1:7
        LE(tt+it-90)=mm*(tt+it)+cc;
    end
end
length(LE(1:163));
length(91:1:253);
%Process Q*R*Q+*x(t)
%Assign omega=g
g=pi/4;
%Creates list of Q+*x(t)
V=zeros(163,1);
for ii=0:32;
QX=zeros(M,1);
for k=-N:1:N
    QX(k+N+1)=0;
    for t=ts;
        QX(k+N+1)=QX(k+N+1)+(g/pi)*sinc((k*pi+g*t)/X)*z(t);
    end
end
%Creates matrix Rkm
R=zeros(M,M);
for k=-N:1:N
    for m=-N:1:N
        for t=ts

R(k+N+1,m+N+1)=R(k+N+1,m+N+1)+(g/pi)^2*sinc((k*pi+g*t)/X)*sinc((m*pi+g*t)/X
);
        end
    end
end
%where v is epsilon
v=0.1;
RI= R+eye(M)*v;
y=inv(RI)*QX;
%Creates list of Q sum of all Yk for each t
Q=zeros(length(ts)+9,1);
for t=q:1:(s+9)
    for k=-N:1:N
        Q(t-ii*5)=Q(t-ii*5)+y(k+N+1)*(g/pi)*sinc((k*pi+g*t)/X);
    end
end
%Calulating residual
t=q:1:s+9;
```

```matlab
BLR=0;
D=zeros(length(Q),1);
D=Q(:);
%for i=1:1:length(Q)
    %for j=1:1:12
        %if zm(i)==j
         %   D(i)=D(i)+S(j);
         %       end

    % end
%end
for i=1:1:91
    BLR=BLR+sqrt((z(i+ii*5)-(D(i)))^2);
end
BLR
BLR/90
D(1:91)=D(1:91)+(BLR/s);
D(92:100)=D(92:100)+((MV(87+5*ii)+MV(88+5*ii)+MV(89+5*ii)+MV(90+5*ii)+MV(86
+5*ii))/5);
V(q:q+8)=D(92:100);
s=s+5;
q=q+5;
time=time+1
end
BLRE=0;
V;
for i=1:1:162
    BLRE=BLRE+sqrt((z(i+91)-(V(i)))^2);
end
LERR=0;
for i=1:1:162
    LERR=LERR+sqrt((z(i+91)-(LE(i)))^2);
end
length(V);
BLRE
BLRE/length(V(1:162))
LERR
LERR/length(V(1:162))
length(V(1:160));
length(z(91:250));
length(MV(91:250));
tt=92:1:251;
length(tt);

%Extrapolation
hold on
plot(1:1:90,z(1:90),'k.:',1:1:90,MV(1:90),'r','HandleVisibility','off')
plot(tt,z(91:250),'k.:',tt,V(1:160),'b--',tt,MV(91:250),'r')
title('Left band limited extrapolation of ASX stock price, 5 day forecast
for 160 days ')
xlabel('Days')
ylabel('Max price $')
line([90,90],[40,100],'Color',[0,0,0],'HandleVisibility','off')
hold all
plot(91:1:253,LE(1:163),'g--')
legend('Raw data','Band limited extrapolation','Five point moving
average','Linear extrapolation')
%%Band limited extrapolation  residual   254.1679
%%Band limited extrapolation average residual   1.5689
%%Linear extrapolation residual 357.9508
%%Linear extrapolation average residual 2.2096
```

# Appendix 8.5 Band Limited Extrapolation of ASX: CBA Data 2 Day forecasts

```matlab
%Start by importing raw data
A=xlsread('C:\Users\valough\Documents\MATLAB\CBA.AX.csv');
z=transpose(A(:,5));
zm=transpose(A(:,3));
N=45;
n=-N:1:N;
M=2*N+1;
X=pi;
%Currently unused
s=91;
q=1;
ts=q:1:s;
time=0
%Moving average
MV=zeros(length(z),1);
MV(1)=(z(1)+z(2)+z(3)+z(4)+z(5))/5;
MV(1)=MV(2);
MV(1)=MV(3);
MV(length(z)-1)=(z(length(z)-1)+z(length(z)-2)+z(length(z)-3)+z(length(z)-4)+z(length(z)))/5;
MV(length(z)-1)=MV(length(z)-2);
MV(length(z)-1)=MV(length(z)-3);
for l=3:(length(z)-3)
    MV(l)=(z(l)+z(l+1)+z(l+2)+z(l-2)+z(l-1))/5;
end
MA=mean(MV(5:length(MV)-5));
LE=zeros(10,1);
for tt=91:2:253
    mm=(z(tt)-z(tt-1))/2;
    cc=(z(tt)-((z(tt)-z(tt-1))/2)*tt);
    for it=1:2
        LE(tt+it-90)=mm*(tt+it)+cc;
    end
end
length(LE(1:163))
length(91:1:253)
%Process Q*R*Q+*x(t)
%Assign omega=g
g=pi/4;
%Creates list of Q+*x(t)
V=zeros(25,1);
for ii=0:79;
QX=zeros(M,1);
for k=-N:1:N
    QX(k+N+1)=0;
    for t=ts;
        QX(k+N+1)=QX(k+N+1)+(g/pi)*sinc((k*pi+g*t)/X)*MV(t);
    end
end
%Creates matrix Rkm
R=zeros(M,M);
for k=-N:1:N
    for m=-N:1:N
        for t=ts
```

```matlab
            R(k+N+1,m+N+1)=R(k+N+1,m+N+1)+(g/pi)^2*sinc((k*pi+g*t)/X)*sinc((m*pi+g*t)/X);
        end
    end
end
%where v is epsilon
v=0.1;
RI= R+eye(M)*v;
y=inv(RI)*QX;
%Creates list of Q sum of all Yk for each t
Q=zeros(length(ts)+14,1);
for t=q:1:(s+6)
    for k=-N:1:N
        Q(t-ii*2)=Q(t-ii*2)+y(k+N+1)*(g/pi)*sinc((k*pi+g*t)/X);
    end
end
%Calulating residual
t=q:1:s+6;
BLR=0;
D=zeros(length(Q),1);
D=Q(:);
%for i=1:1:length(Q)
    %for j=1:1:12
        %if zm(i)==j
        %   D(i)=D(i)+S(j);
        %      end

    % end
%end
for i=1:1:91
    BLR=BLR+sqrt((z(i+ii*2)-(D(i)))^2);
end
BLR
BLR/90
D(1:91)=D(1:91)+(BLR/s);
D(92:98)=D(92:98)+(MV(87+2*ii)+MV(88+2*ii)+MV(89+2*ii)+MV(90+2*ii)+MV(91+2*ii))/5;
V(q:q+6)=D(92:98);
s=s+2;
q=q+2;
time=time+1
end
BLRE=0;

for i=15:1:162
    BLRE=BLRE+sqrt((z(i+91)-(V(i)))^2);
end
LERR=0;
for i=15:1:162
    LERR=LERR+sqrt((z(i+91)-(LE(i)))^2);
end
length(V);
BLRE
BLRE/length(V(1:162))
LERR
LERR/length(V(1:162))
length(LE(1:163))
length(V(1:163));
length(z(91:253));
length(MV(91:253));
```

```
tt=91:1:253;
length(tt);

%Extrapolation

%Extrapolation
hold on
plot(1:1:90,z(1:90),'k.:',1:1:90,MV(1:90),'r','HandleVisibility','off')
plot(tt,z(91:253),'k.:',tt,V(1:163),'b--',tt,MV(91:253),'r')
title('Left band limited extrapolation of ASX stock price, 2 day forecast
for 160 days')
xlabel('Days')
ylabel('Max price $')
line([90,90],[40,100],'Color',[0,0,0],'HandleVisibility','off')
hold all
plot(91:1:253,LE(1:163),'g--')
legend('Raw data','Band limited extrapolation','Five point moving
average','Linear extrapolation')
%%Band limited extrapolation  residual    147.8019
%%Band limited extrapolation average residual    0.9987
%%Linear extrapolation residual 142.0342
%%Linear extrapolation average residual 0.9532
```

# Appendix 8.6  Band Limited Extrapolation of BOM:2017 Maximum temperature 14 day forecast

```
%Start by importing raw data
A=xlsread('C:\Users\valough\Documents\MATLAB\IDCJAC0010_009225_2017_Data.cs
v');
z=transpose(A(:,5));
zm=transpose(A(:,3));
B=xlsread('C:\Users\valough\Documents\MATLAB\IDCJAC0010_009225_2016_Data.cs
v');
Bm=transpose(B(:,3));
Bz=transpose(B(:,5));
X=pi;
N=45;
n=-N:1:N;
M=2*N+1;
%Currently unused
s=0;
q=-90;
ts=q:1:s;
BLRE=0;
time=0

%Moving average
MV=zeros(length(z),1);
MV(1)=(z(1)+z(2)+z(3)+z(4)+z(5))/5;
MV(1)=MV(2);
MV(1)=MV(3);
MV(length(z)-1)=(z(length(z)-1)+z(length(z)-2)+z(length(z)-3)+z(length(z)-
4)+z(length(z)))/5;
MV(length(z)-1)=MV(length(z)-2);
MV(length(z)-1)=MV(length(z)-3);
for l=3:(length(z)-3)
    MV(l)=(z(l)+z(l+1)+z(l+2)+z(l-2)+z(l-1))/5;
end
MA=mean(MV(5:length(MV)-5));
%Process Q*R*Q+*x(t)
```

```matlab
%Assign omega=g
g=pi/4;
%Creates list of Q+*x(t)
QX=zeros(M,1);
for k=-N:1:N
    QX(k+N+1)=0;
    for t=ts;
        QX(k+N+1)=QX(k+N+1)+(g/pi)*sinc((k*pi+g*t)/X)*MV(t+91);
    end
end
%Creates matrix Rkm
R=zeros(M,M);
for k=-N:1:N
    for m=-N:1:N
        for t=ts

R(k+N+1,m+N+1)=R(k+N+1,m+N+1)+(g/pi)^2*sinc((k*pi+g*t)/X)*sinc((m*pi+g*t)/X
);
        end
    end
end
%where v is epsilon
v=0.1;
RI= R+eye(M)*v;
y=inv(RI)*QX;
%Creates list of Q sum of all Yk for each t
Q=zeros(length(ts)+14,1);
for t=q:1:(s+14)
    for k=-N:1:N
        Q(t+91)=Q(t+91)+y(k+N+1)*(g/pi)*sinc((k*pi+g*t)/X);
    end
end
%Calulating residual
t=q:1:s+14;
BLR=0;

D=Q(:);
%for i=1:1:length(Q)
    %for j=1:1:12
        %if zm(i)==j
        %   D(i)=D(i)+S(j);
        %       end

    % end
%end
for i=1:1:90
    BLR=BLR+sqrt((z(i)-(D(i)))^2);
end
MA;
BLR
BLR/89
mean(Q)
mean(z)
D(1:91)=D(1:91)+BLR/90;
ii=0
for td = 2:5

D(90+td)=D(90+td)+(MV(86+7*ii)+MV(87+7*ii)+MV(88+7*ii)+MV(89+7*ii)+MV(90+7*
ii))/5-((4-td)*BLR)/(5*90);
end
```

```matlab
        D(96:105)=D(96:105)+(MV(86+7*ii)+MV(87+7*ii)+MV(88+7*ii)+MV(89+7*ii)+MV(90+7*ii))/5;
% D(90:95)
 % 19.2264
 %  16.3732
  % 11.8949
   %  6.6429
    % 1.7988
   % -1.5683
  BLR=0;
  for i=1:1:90
    BLR=BLR+sqrt((z(i)-(D(i)))^2);
  end
BLR
for i=91:1:105
    BLRE=BLRE+sqrt((z(i)-(D(i)))^2);
end
BLRE
length(z(15:119))
length(MV(15:119))
length(t)
length(D)
%Extrapolation
plot(t,z(1:105),'k.:',t,D,'b--',t,MV(1:105),'r')
title('Left band limited extrapolation of the Maximum temperature at Perth Metro, single test 90 days of data')
legend('Raw data','Band limited extrapolation','Five point moving average')
xlabel('Days')
ylabel('Max Temperature C')
line([0,0],[min(z),max(z)],'Color',[0,0,0])
```

## Appendix 8.8 Band Limited Extrapolation of BOM:2017 Maximum temperature Data 5 Day forecasts

```matlab
%Start by importing raw data
A=xlsread('C:\Users\valough\Documents\MATLAB\IDCJAC0010_009225_2017_Data.csv');
z=transpose(A(:,5));
zm=transpose(A(:,3));
N=45;
n=-N:1:N;
M=2*N+1;
%Currently unused
s=0;
q=-90;
ts=q:1:s;
time=0
X=pi;
%Moving average
MV=zeros(length(z),1);
MV(1)=(z(1)+z(2)+z(3)+z(4)+z(5))/5;
MV(1)=MV(2);
MV(1)=MV(3);
MV(length(z)-1)=(z(length(z)-1)+z(length(z)-2)+z(length(z)-3)+z(length(z)-4)+z(length(z)))/5;
MV(length(z)-1)=MV(length(z)-2);
MV(length(z)-1)=MV(length(z)-3);
for l=3:(length(z)-3)
```

```matlab
        MV(l)=(z(l)+z(l+1)+z(l+2)+z(l-2)+z(l-1))/5;
end
MA=mean(MV(5:length(MV)-5));
%Process Q*R*Q+*x(t)
%Assign omega=g
g=pi/4;
%Creates list of Q+*x(t)
V=zeros(264,1);
for ii=0:38;
QX=zeros(M,1);
for k=-N:1:N
    QX(k+N+1)=0;
    for t=ts;
        QX(k+N+1)=QX(k+N+1)+(g/pi)*sinc((k*pi+g*t)/X)*MV(t+91+ii*7);
    end
end
%Creates matrix Rkm
R=zeros(M,M);
for k=-N:1:N
    for m=-N:1:N
        for t=ts

R(k+N+1,m+N+1)=R(k+N+1,m+N+1)+(g/pi)^2*sinc((k*pi+g*t)/X)*sinc((m*pi+g*t)/X);
        end
    end
end
%where v is epsilon
v=0.1;
RI= R+eye(M)*v;
y=inv(RI)*QX;
%Creates list of Q sum of all Yk for each t
Q=zeros(length(ts)+14,1);
for t=q:1:(s+14)
    for k=-N:1:N
        Q(t+91)=Q(t+91)+y(k+N+1)*(g/pi)*sinc((k*pi+g*t)/X);
    end
end
%Calulating residual
t=q:1:s+14;
BLR=0;
D=zeros(length(Q),1);
D=Q(:);
%for i=1:1:length(Q)
    %for j=1:1:12
        %if zm(i)==j
         %   D(i)=D(i)+S(j);
         %       end

    % end
%end
for i=1:1:91
    BLR=BLR+sqrt((z(i+ii*7)-(D(i)))^2);
end
BLR
BLR/90
D(1:91)=D(1:91)+(BLR/90);
D(92:105)=D(92:105)+MV(90+7*ii);
%D(92:105)=D(92:105)+(MV(87+7*ii)+MV(88+7*ii)+MV(89+7*ii)+MV(90+7*ii)+MV(91+7*ii))/5;
V(q+91+ii*7:q+102+ii*7)=D(94:105);
```

```
time=time+1
end
BLRE=0;
V;
for i=1:1:267
    BLRE=BLRE+sqrt((z(i+91)-(V(i)))^2);
end
length(D);
BLRE
BLRE/length(V)
length(V(1:267))
length(z(91:357))
length(MV(91:357))
tt=92:1:358;
length(tt)

%Extrapolation
hold on
plot(1:1:90,z(1:90),'k.:',1:1:90,MV(1:90),'r','HandleVisibility','off')
plot(tt,z(92:358),'k.:',tt,V(1:267),'b--',tt,MV(92:358),'r')
title('Left band limited extrapolation of the Maximum temperature at Perth
Metro, 38 weeks of tests, 90 days of data moving')
xlabel('Days')
ylabel('Temperature C')
line([91,91],[0,40],'Color',[0,0,0],'HandleVisibility','off')
legend('Raw data','Band limited extrapolation','Five point moving average')
hold off
%BLRE 587.4170   2.1837
```

## Appendix 8.9 Band Limited Extrapolation of BOM:2017 Maximum temperature Data 2 Day forecasts

```
%Start by importing raw data
A=xlsread('C:\Users\valough\Documents\MATLAB\IDCJAC0010_009225_2017_Data.cs
v');
z=transpose(A(:,5));
zm=transpose(A(:,3));
N=45;
n=-N:1:N;
M=2*N+1;
X=pi;
%Currently unused
s=91;
q=1;
ts=q:1:s;
time=0
%Moving average
MV=zeros(length(z),1);
MV(1)=(z(1)+z(2)+z(3)+z(4)+z(5))/5;
MV(1)=MV(2);
MV(1)=MV(3);
MV(length(z)-1)=(z(length(z)-1)+z(length(z)-2)+z(length(z)-3)+z(length(z)-
4)+z(length(z)))/5;
MV(length(z)-1)=MV(length(z)-2);
MV(length(z)-1)=MV(length(z)-3);
for l=3:(length(z)-3)
    MV(l)=(z(l)+z(l+1)+z(l+2)+z(l-2)+z(l-1))/5;
```

```matlab
end
MA=mean(MV(5:length(MV)-5));
%Process Q*R*Q+*x(t)
%Assign omega=g
g=pi/4;
%Creates list of Q+*x(t)
V=zeros(25,1);
for ii=0:133;
QX=zeros(M,1);
for k=-N:1:N
    QX(k+N+1)=0;
    for t=ts;
        QX(k+N+1)=QX(k+N+1)+(g/pi)*sinc((k*pi+g*t)/X)*z(t);
    end
end
%Creates matrix Rkm
R=zeros(M,M);
for k=-N:1:N
    for m=-N:1:N
        for t=ts

R(k+N+1,m+N+1)=R(k+N+1,m+N+1)+(g/pi)^2*sinc((k*pi+g*t)/X)*sinc((m*pi+g*t)/X);
        end
    end
end
%where v is epsilon
v=0.1;
RI= R+eye(M)*v;
y=inv(RI)*QX;
%Creates list of Q sum of all Yk for each t
Q=zeros(length(ts)+14,1);
for t=q:1:(s+6)
    for k=-N:1:N
        Q(t-ii*2)=Q(t-ii*2)+y(k+N+1)*(g/pi)*sinc((k*pi+g*t)/X);
    end
end
%Calulating residual
t=q:1:s+6;
BLR=0;
D=zeros(length(Q),1);
D=Q(:);
%for i=1:1:length(Q)
    %for j=1:1:12
         %if zm(i)==j
          %    D(i)=D(i)+S(j);
          %        end

    % end
%end
for i=1:1:91
    BLR=BLR+sqrt((z(i+ii*2)-(D(i)))^2);
end
BLR
BLR/91
D(1:91)=D(1:91)+(BLR/s);
D(92:98)=D(92:98)+(z(87+2*ii)+z(88+2*ii)+z(89+2*ii)+z(90+2*ii)+z(91+2*ii))/5;
V(q:q+6)=D(92:98);
s=s+2;
q=q+2;
```

```matlab
    time=time+1
end
BLRE=0;
V
for i=1:1:250
    BLRE=BLRE+sqrt((z(i+91)-(V(i+2)))^2);
end
length(V)
BLRE
BLRE/length(V(3:257))
length(V(1:255))
length(z(91:345))
length(MV(91:345))
tt=91:1:345;
length(tt)

%Extrapolation
hold on
plot(1:1:90,z(1:90),'k.:',1:1:90,MV(1:90),'r','HandleVisibility','off')
plot(tt,z(91:345),'k.:',tt,V(3:257),'b--',tt,MV(91:345),'r')
title('Left band limited extrapolation of the Maximum temperature at Perth Metro, 38 weeks of tests, 90 days of data moving')
xlabel('Days')
ylabel('Temperature C')
line([91,91],[0,40],'Color',[0,0,0],'HandleVisibility','off')
legend('Raw data','Band limited extrapolation','Five point moving average')
hold off
```